# Einstein Gravitation Theory: Experimental Tests II

M. Cattani
Instituto de Fisica, Universidade de S. Paulo, C.P. 66318, CEP 05315–970
S. Paulo, S.P. Brazil . E–mail: mcattani@if.usp.br

Abstract.
In order to test the Einstein gravitation theory (EGT) we compare their predictions with the measured results in the following phenomena: the perihelion advance of planets, deflection of light, radar echo delays around the Sun and an overall planetary motion in Solar System. In our calculation we have used the Schwarzschild metric that is defined in the surrounding vacuum of a spherically symmetric mass distribution, not in rotation. This article was written to graduate and postgraduate students of Physics.

Resumo.
Com intuito de testar a validade da teoria de gravitação de Einstein (TGE) comparamos as suas previsões com os resultados medidos nos seguintes fenômenos: precessão do periélio de planetas, deflexão da luz, atraso temporal de ecos de sinais de radar ao redor do Sol e movimentos planetários no Sistema Solar. Nesses cálculos usamos a métrica de Schwarzschild que é definida no vácuo em torno de uma distribuição esfericamente simétrica de massa, não em rotação. Esse artigo foi escrito para alunos de graduação e pós–graduação de Física.

## I. Introdução.

Graças aos grandes avanços tecnológicos que surgiram a partir de 1960 vários testes muito precisos foram e ainda estão sendo feitos afim de confirmar a Teoria de Gravitação de Einstein (TGE). Nesses experimentos usam–se sondas espaciais, relógios atômicos, eletrônica ultra–rápida, detectores ultra–sensíveis de luz (CCD) e de partículas, radares, super–computadores, lasers de alta potência, observatórios espaciais e terrestres, etc. Com essa parafernália tecnológica uma quantidade enorme de dados foram e estão sendo coletados de todo o Universo. Eles estão sendo analisados à luz das previsões das teorias relativísticas vigentes.[1,2]

No artigo precedente[3] usando a TGE calculamos a métrica do espaço–tempo denominada de métrica de Schwarzschild que é gerada no vácuo ao redor de uma distribuição esfericamente simétrica de massa, não em rotação. Há muitos fenômenos nessas condições onde a TGE pode ser testada, não só no sistema solar[1,2,4–8] mas também em escala cósmica, tais como dilação temporal, deflexão e efeito Doppler da luz, precessão do periélio de planetas, desvios da teoria Newtoniana nos movimentos planetários, precessão geodética de giroscópios, atraso temporal de ecos de sinais de radar passando ao redor do Sol, lentes gravitacionais, buracos negros e emissão de ondas gravitacionais[9] pelo binário constituído pelo pulsar PSR 1913 + 16.

De acordo com o artigo anterior[3] a métrica de Schwarzschild é definida através do invariante $ds^2$ dado por

$$ds^2 = (1- 2GM/c^2r)\, c^2dt^2 - dr^2/(1 - 2GM/c^2r) - r^2d\theta^2 - r^2\sin^2\theta\, d\varphi^2 \quad (I.1)$$

ou por,

$$ds^2 = [(1- 2GM/c^2r)/(1+2GM/c^2r)]^2\, c^2dt^2 -$$

$$(1+2GM/c^2r)^4\, (dr^2 + r^2d\theta^2 + r^2\sin^2\theta\, d\varphi^2) \quad (I.2).$$

O elemento de linha dado pela (I.2) é conhecido como *elemento de linha isotrópico*. Adotando essa métrica[3] analisamos a *dilação temporal* e o *efeito Doppler da luz*. Agora vamos analisar a *precessão do periélio dos planetas*, a *deflexão de luz pelo Sol*, o *retardo de ecos de sinais de radar passando ao redor do Sol* e o *movimento planetário no Sistema Solar*.

Seguindo o procedimento adotado nos artigos precedentes[9] procuraremos citar um mínimo possível de referências (artigos e livros) fazendo os cálculos com suficiente precisão, deixando de lado alguns refinamentos que poderão ser encontrados em artigos. As previsões serão comparadas com resultados experimentais sem a preocupação exagerada de analisar detalhes das técnicas empregadas e de suas limitações. Esses aspectos poderão ser vistos nas referências citadas.

## 1. Precessão do Periélio de Planetas.

Para estudarmos o movimento planetário adotamos a métrica dada por (I.2) que pode ser escrita na forma

$$ds^2 = e^{-2\alpha}\, c^2dt^2 - e^{2\beta}(dr^2 + r^2d\theta^2 + r^2\sin^2\theta\, d\varphi^2) \quad (1.1),$$

onde  $e^{-2\alpha} = g_{oo}(r) = g_{44}(r) = (1- 2GM/c^2r)/(1+2GM/c^2r)$,

$e^{2\beta} = g_{11}(r) = (1+2GM/c^2r)^4$,   $g_{22}(r) = r^2 e^{2\beta}$   e   $g_{33}(4) = r^2\sin^2\theta\, e^{2\beta}$,

levando em conta que $x_o = x_4 = ct$, $x_1 = r$, $x_2 = \theta$ e $x_3 = \varphi$. Com esses valores podemos calcular os símbolos de Christoffel definidos por [4–8,10,11]

$$\Gamma_{\mu\nu}{}^{\alpha} = \{{}_{\mu}{}^{\alpha}{}_{\nu}\} = (g^{\alpha\lambda}/2)(\partial_{\nu} g_{\lambda\mu} + \partial_{\mu} g_{\lambda\nu} - \partial_{\lambda} g_{\mu\nu}) \qquad (1.2),$$

lembrando[4] que $g^{\mu\nu} = M_{\mu\nu}/|g|$ onde g é o determinante de $g_{\mu\nu}$ e $M_{\mu\nu}$ é o determinante menor de $g_{\mu\nu}$ em g. Como os elementos de g são diagonais temos $|g| = |g_{11} g_{22} g_{33} g_{44}| = c^2 e^{2(\beta - \alpha)} r^4 \sin^2\theta$. Assim, $g^{oo} = c^{-2} e^{2\alpha}$, $g^{11} = e^{-2\beta}$, $g^{22} = r^{-2} e^{-2\beta}$ e $g^{33} = r^{-2} \sin^{-2}\theta\, e^{-2\beta}$. Como os $g_{\mu\nu}$ só dependem de $r = x_1$ em (2.2) só há derivadas do tipo $\partial_r g_{\mu\nu} = \partial_1 g_{\mu\nu}$. Indicando por $\alpha' = \partial\alpha/\partial r$ e $\beta' = \partial\beta/\partial r$ obtemos

$\Gamma_{o1}{}^{o} = \{{}_{o}{}^{o}{}_{1}\} = \Gamma_{1o}{}^{o} = \{{}_{1}{}^{o}{}_{o}\} = -\alpha',$  $\qquad \Gamma_{oo}{}^{1} = \{{}_{o}{}^{1}{}_{o}\} = \alpha' e^{-2(\alpha+\beta)},$

$\Gamma_{11}{}^{1} = \{{}_{1}{}^{1}{}_{1}\} = \beta'$ $\qquad \Gamma_{23}{}^{3} = \{{}_{2}{}^{3}{}_{3}\} = \Gamma_{32}{}^{3} = \{{}_{3}{}^{3}{}_{2}\} = \cot\theta,$

$\Gamma_{12}{}^{2} = \{{}_{1}{}^{2}{}_{2}\} = \Gamma_{21}{}^{2} = \{{}_{1}{}^{3}{}_{3}\} = 1/r + \beta',$ $\quad \Gamma_{22}{}^{1} = \{{}_{2}{}^{1}{}_{2}\} = r + r^2 \beta',$ $\qquad (1.3)$

$\Gamma_{13}{}^{3} = \{{}_{1}{}^{3}{}_{3}\} = \Gamma_{31}{}^{3} = \{{}_{3}{}^{3}{}_{1}\} = 1/r + \beta',$ $\quad \Gamma_{33}{}^{2} = \{{}_{3}{}^{2}{}_{3}\} = -\sin\theta \cos\theta$ e

$\Gamma_{33}{}^{1} = \{{}_{3}{}^{1}{}_{3}\} = -(r + r^2 \beta') \sin^2\theta$

Assumindo que o planeta se move em um plano (plano equatorial xy) poremos $\theta$ = constante = $\pi/2$ em (1.3). Nesse caso (1.1) fica escrita como

$$ds^2 = e^{-2\alpha} c^2 dt^2 - e^{2\beta}(dr^2 + r^2 d\varphi^2) \qquad (1.4).$$

O planeta percorre uma geodésica que obedece a equação [4–8,10,11]

$$d^2x^{\alpha}/ds^2 + \Gamma_{\tau\nu}{}^{\alpha} (dx^{\nu}/ds)(dx^{\tau}/ds) = 0 \qquad (1.5).$$

Substituindo em (2.5) os $\Gamma_{\tau\nu}{}^{\alpha}$ dados pela (1.3) com $\theta$ = constante = $\pi/2$ teremos, (*)

$d^2r/ds^2 + \beta'(dr/ds)^2 - 2(r + r^2\beta')(d\varphi/ds)^2 - \alpha' e^{-2(\alpha+\beta)}(dt/ds)^2 = 0$

$d^2\varphi/ds^2 + 2(1/r + \beta')(d\varphi/ds)(dr/ds) = 0 \qquad (1.6)$

$d^2t/ds^2 - 2\alpha'(dt/ds)(dr/ds) = 0.$

Pegando a segunda equação de (1.6) e pondo $\eta = d\varphi/ds$ obtemos,

$d\eta/ds + 2(1/r + d\beta/dr)\eta (dr/ds) = 0$  ou  $d\eta/\eta = -2(1/r + d\beta/dr)dr,$

que integrando dá

$$d\varphi/ds = A\, e^{-2\beta}/r^2 \qquad (1.7),$$

onde A é uma constante. Analogamente, pegando a terceira equação de (1.6) e pondo $\psi = dt/ds$ obtemos, sendo B = constante:

$$dt/ds = B\, e^{2\alpha} \qquad (1.8),$$

Levando em conta que[11] $ds/dt = c\, [1-(v/c)^2]^{1/2}$ e que de acordo com (1.1), $e^{2\beta} = (1+2GM/c^2r)^4$ verificamos que para o limite não relativístico, ou seja, quando $e^{2\beta} \to 1$ e $ds \to c\, dt$, de (1.7) obtemos $r^2 d\varphi/dt = cA = $ constante que corresponde à lei das áreas de Kepler de onde resulta a lei de conservação do momento angular $L_\theta = m\, r^2 (d\varphi/dt)$ do planeta de massa m. Assim vemos que devemos ter $mcA = L_\theta$. De modo análogo pode−se mostrar[4] que a (1.8) no limite não relativístico corresponde à conservação da energia.

Para calcular a órbita (trajetória) do planeta vamos substituir dt dada por (1.8), $dt = B\, e^{2\alpha}$ e $ds = r^2 e^{2\beta}\, d\varphi/A$, dada por (1.7), no elemento de linha ds definido por (1.4) obtendo,

$$(dr/d\varphi)^2 + r^2 = (r^4/A^2)\, e^{2\beta}\, (B^2 e^{2\alpha} - 1) \qquad (1.9).$$

Como $G/c^2 = 7.414\; 10^{-28}$ m/kg, a massa $M_S$ do Sol é $M_S = 2.3\; 10^{30}$ kg e a distância r dos planetas ao Sol é $r \geq 5.8\; 10^{10}$ m verificamos que o parâmetro $GM_S/c^2 r \ll 10^{-8}$. Nessas condições expandindo $e^{2\alpha}$ e $e^{2\beta}$ em termos de $\kappa = GM_S/c^2$ e considerando somente termos até segunda ordem em $\kappa/r$ obtemos,

$$e^{2\alpha} = 1 + 2\kappa/r + 2\kappa^2/r^2 + ....$$

$$e^{2\beta} = 1 + 2\kappa/r + 3\kappa^2/2r^2 + ...$$

$$B = e^{-2\alpha}\, (dt/ds) \approx 1/c \qquad (1.10)$$

$$(dr/d\varphi)^2 + r^2 = (r^4/A^2)\, (2\kappa/r + 6\kappa^2/r^2).$$

Pondo $u = 1/r$ a ultima equação de (1.10) ela fica escrita como,

$$(du/d\varphi)^2 + u^2 = (1/A^2)\, (2\kappa u + 6\kappa^2 u^2) \qquad (1.11).$$

Diferenciando a (1.11) em relação a $\varphi$ obtemos finalmente,

______________________________________________________________

(*) Se não assumíssemos que $\theta = $ constante $= \pi/2$ apareceria em (1.6) uma equação a mais do tipo[8] $d^2\theta/dt^2 + 2(dr/dt)(d\theta/dt) - \cos\theta \sin\theta\, (d\varphi/dt)^2 = 0$. Assim, se inicialmente $\theta = \pi/2$ e $(d\theta/dt) = 0$, teríamos $d^2\theta/dt^2 = 0$ mostrando que a órbita permaneceria constantemente no plano (x,y).

$$d^2u/d\varphi^2 + u = \kappa/A^2 + (6\kappa^2/A^2)u \qquad (1.12).$$

No caso do movimento planetário não−relativístico, levando em conta que $A \to L_\theta/mc$ e $\kappa = GM/c^2$ ao invés da (1.12) temos [5,12–14]

$$d^2u/d\varphi^2 + u = GM_S m^2/L_\theta^{\,2} = GM_S/C^2 \qquad (1.13),$$

onde $L_\theta = m\, r^2\, (d\varphi/dt) = mC$ e $C/2 = r^2\, (d\varphi/dt)/2 =$ constante é a velocidade areolar do planeta. A (1.13) tem como solução a função [5,12–14]

$$u = 1/r = (GM_S/C^2)\,(1 + \varepsilon\cos\varphi) \qquad (1.14),$$

que é a equação de uma elipse com a origem em um dos focos (onde está o Sol) onde $\varepsilon = b/(a^2 - b^2)$ é a sua ecentricidade, a o raio maior e b o raio menor. No periélio ($\varphi = 0$) a distância do planeta ao Sol é dada por $r_{min} = (C^2/GM_S)/(1+\varepsilon)$, o planeta está mais perto do Sol. No afélio ($\varphi = \pi$), $r_{max} = (C^2/GM_S)/(1-\varepsilon)$. A elipse definida por (1.14) é dita "Newtoniana".

Não levando em conta o termo $(6\kappa^2/A^2)u$ na (1.12) ficaríamos com

$$d^2u/d\varphi^2 + u = \kappa/A^2 \qquad (1.15).$$

que integrada daria uma elipse $u = (\kappa/A^2)(1 + \varepsilon'\cos\varphi)$ com parâmetros praticamente idênticos aos obtidos no caso Newtoniano. Vamos agora levar em conta o termo $(6\kappa^2/A^2)u$ e colocar (1.12) na forma

$$d^2u/d\varphi^2 + (1 - 6\kappa^2/A^2)u = \kappa/A^2 \qquad (1.15),$$

que tem como solução, desprezando termos da ordem de $(\kappa/r)^3$, a função

$$u \approx (\kappa/A^2)\,(1 + \varepsilon'\cos\rho\varphi) \qquad (1.16),$$

onde $\rho = (1 - 6\kappa^2/A^2)^{1/2}$. Como $\kappa^2/A^2 = GM_S/Rc^2 \sim 10^{-8}$, conforme resultados numéricos vistos acima, podemos fazer $\rho = (1 - 6\kappa^2/A^2)^{1/2} \approx 1 - 3\kappa^2/A^2$ e $\varepsilon' \approx \varepsilon$ definida em (1.14). O fator $\rho$ gera uma precessão do periélio[4–8]. De fato, sejam $\varphi_1$ e $\varphi_2$ dois ângulos adjacentes no periélio para os quais temos $(1 - 3\kappa^2/A^2)\varphi_2 = 2\pi + (1 - 3\kappa^2/A^2)\varphi_1$. Desse modo vemos que $\Delta\varphi = \varphi_2 - \varphi_1$ é dado por

$$\Delta\varphi = 2\pi/(1 - 3\kappa^2/A^2) \approx 2\pi(1 + 3\kappa^2/A^2) \qquad (1.17),$$

Mostrando que a distância angular entre um periélio e o seguinte é maior do que $2\pi$ por um fator

$$6\pi\,\kappa^2/A^2 = 6\pi\,(GM_S/c^2 A)^2 \qquad (1.18).$$

Como a órbita é aproximadamente circular com raio R vemos que $\kappa/A^2 = (GM_S/c^2)/(C/c)^2 \approx 1/R$. Assim, o *avanço angular do periélio por revolução* é dado por

$$6\pi\, GM_S/Rc^2 \qquad (1.19).$$

Na Figura 1 mostramos o avanço angular $\Delta\varphi$ com o auxílio da trajetória descrita pelo planeta em torno do Sol (no centro da figura) que ocupa um dos focos da elípse precessionante.

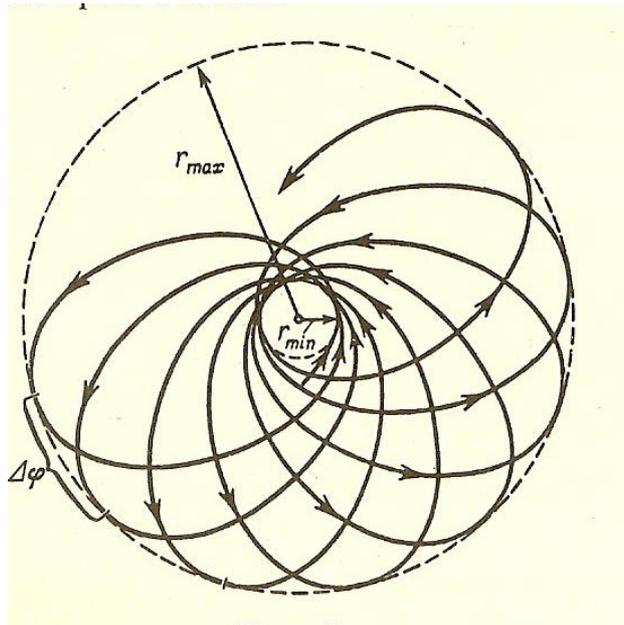

**Figura 1.** Ilustração do avanço angular $\Delta\varphi$ do periélio de um planeta.[10] O $r_{min}$ é o periélio e $r_{max}$ é o afélio.

Para Mercúrio[2,7,8] temos $\Delta\varphi \sim 0.10"$ por revolução ou $\Delta\varphi \sim 40"$ *por século*. Levando em conta ecentricidade da órbita o raio R é substituído por $(r_{max} + r_{min})/2$. Com essa correção temos os seguintes resultados teóricos e experimentais,[2,7,8] respectivamente, para a precessão por século: 43.03" e $(43.11 \pm 0.45)"$ para Mercúrio, 8.6" e $(8.4 \pm 4.8)"$ para Vênus e 3.8" e $(5.0 \pm 1.2)"$ para a Terra. Assim, podemos concluir que a TGE descreve muito bem as precessões observadas dos periélios de planetas.

Cálculos mais precisos para Mercúrio levando em conta a sua massa e a do Sol, ecentricidade da órbita, quadrupolo do Sol e outros fatores são mencionados na Seção (3.5) do artigo e livro de Will[1,2].

## 2. Deflexão da Luz.

A nossa meta é calcular a deflexão luminosa com pequeno ângulo que se observa, por exemplo, quando a luz emitida por estrelas ou quasares é desviada pelo Sol. Estudos sobre "lentes gravitacionais" e os efeitos gravitacionais intensos gerados nas vizinhanças de buracos negros podem ser vistos em outros artigos e livros[1,2] (vide comentários na Seção 2.1).

Como a luz descreve uma geodésica de comprimento nulo para calcularmos a sua trajetória temos de assumir que ds = 0 em (1.4)–(1.7). De (1.7) resulta que A → ∞ e B → ∞ e, consequentemente, que

$$d\varphi/dt = \lambda e^{-2(\alpha+\beta)}/r^2 \qquad (2.1),$$

onde $\lambda$ = A/B é uma constante finita pois $r^2$ (d$\varphi$/dt)/2 = $C$ é a velocidade areolar no limite Newtoniano. Assim, vemos que (1.9) fica dada por

$$(dr/d\varphi)^2 + r^2 = (r^4/\lambda^2) e^{2(\alpha+\beta)} \qquad (2.2).$$

Levando em conta as expansões em série (1.10) de $e^{2\alpha}$ e $e^{2\beta}$, pondo u = 1/r e diferenciando (2.2) em função de $\varphi$ (analogamente ao que foi feito em (1.11)) obtemos, mantendo somente o termo em primeira ordem em κ,

$$d^2u/d\varphi^2 + u = 2\kappa/\lambda^2 \qquad (2.3),$$

que tem como solução a função[5,7,12–14]

$$u = (2\kappa/\lambda^2)(1 + \varepsilon \cos\varphi) \qquad (2.4),$$

que é a equação de uma hipérbole. Diferenciando (2.4) em relação a t obtemos (dr/dt)/$r^2$ = $\varepsilon(2\kappa/\lambda^2)$(d$\varphi$/dt)sin$\varphi$ e lembrando que para r → ∞ temos dr/dt = c e $r^2$(d$\varphi$/dt) = $\lambda$ resulta,

$$\varepsilon = \lambda/2c\,\kappa\,\sin\varphi_\infty \qquad (2.5),$$

onde assumiremos que $\varphi_\infty$ << 1. Como para r → ∞ temos 1/r → 0, de (3.4) verifica-se que cos$\varphi_\infty$ = − 1/$\varepsilon$. Usando este resultado e (2.5) obtemos |tan$\varphi_\infty$| = 2c κ/$\lambda$ ≈ $\varphi_\infty$. Como o ângulo de desvio $\Delta\varphi$ no espalhamento[5,7,12–14] é dado por $\Delta\varphi$ = 2$\varphi_\infty$ deduzimos que $\Delta\varphi$ = 4c κ/$\lambda$. Para um raio de luz proveniente de uma estrela que passa rasante à superfície solar, ou seja, a uma distância r = $R_S$ = raio do Sol, temos $\lambda$ = $R_S$c. Assim, o ângulo de desvio $\Delta\varphi$ fica dado por

$$\Delta\varphi = 4GM_S/R_S c^2 \qquad (2.6).$$

Como $G/c^2 = 7.414 \, 10^{-28}$ m/kg e levando em conta que a massa e o raio do Sol são iguais a $M_S = 1.99 \, 10^{30}$ kg e $R_S = 6.96 \, 10^8$ m, respectivamente, a (2.6) prevê um desvio de 1".75 de arco. No livro de McVittie[4] há uma tabela mostrando vários valores de desvios da luz pelo Sol que foram observados desde 1919(*) até 1952. Essas medidas estavam no limite da precisão dos instrumentos disponíveis na época. Podemos dizer, a grosso modo, que as medidas mais confiáveis estavam no intervalo que vai de 1".72 a 1".82 mostrando um bom acordo com 1".75 previsto pela (2.6). As técnicas usadas nessas medidas foram abandonadas a partir de 1970 por sua pouca precisão e suas inúmeras dificuldades. Outras técnicas muito mais precisas e cômodas começaram a ser desenvolvidas.[1,2] As medidas mais modernas do desvio da luz são feitas atualmente com rádio−telescópios trabalhando na faixa de ondas de rádio o que é válido, pois o desvio $\Delta\varphi$ independe da freqüência da luz.[1,2,15] As fontes de luz preferidas são os quasares. Um quasar ("objeto quase estelar") é uma galáxia jovem e hiper−ativa que está a distâncias enormes (~8 $10^9$ anos−luz) da Terra e que emite uma quantidade fantástica de energia, principalmente na faixa de ondas de rádio. Como está muito longe ele aparece aos radio−telescópios como um ponto luminoso no céu, o que permite uma maior precisão na medida do desvio. Como durante o ano sempre passa um quasar por trás do Sol não é necessário esperar que ocorra um eclipse total do Sol nem se deslocar para sítios distantes onde os eclipses possam ser observados. O raio de luz (ondas de rádio) proveniente do quasar é desviado como se fosse o de uma estrela. Os quasares 3C273 e 3C279 que têm luminosidades intensas já foram usados muitas vezes nas medidas de $\Delta\varphi$ (eles passam por trás do sol no dia 8 de outubro de todo ano). Os valores dessas medidas[2,16,17] concordam com as previsões obtidas com (3.6) com uma precisão melhor que 1% dando suporte à TGE.

2.1 *Lentes gravitacionais*

Outros testes da TGE sobre deflexão da luz estão sendo realizados. Um deles é o efeito de "lente gravitacional" que ocorre quando um quasar passa por trás de uma galáxia próxima (~5 $10^5$ anos−luz da Terra, por exemplo). A luz emitida por ele é detectada na Terra após ser desviada e focalizada pela galáxia que age como uma "lente gravitacional" produzindo múltiplas imagens do quasar.[1,2,7,15] Esse efeito é também investigado observando o céu estelar nas vizinhanças de um buraco negro.[18]

O espectro ótico de um quasar mostra um *redshift* z muito grande $z = (\lambda_o - \lambda_E)/\lambda_E$ onde $\lambda_o$ é o comprimento de onda de uma linha espectral emitida pelo quasar e $\lambda_E$ o comprimento de onda da mesma linha medido na

---------------------------------------------------------------------
(*) As primeiras medidas foram feitas em Sobral (Ceará, Brasil) no dia 29 de maio de 1919.

Terra. O *redshift* z é dado por z = Hd/c onde d = distância do quasar à Terra e H é a constante de Hubble cujo valor medido em 2009 com o auxílio do telescópio Hubble é H = 74.2 ± 3.6 km s$^{-1}$ Mpc$^{-1}$. Os quasares têm valores z ~ 4.5 e localizam–se nos confins do Universo observado. Devido às suas intensas luminosidades acredita–se que sejam núcleos de galáxias jovens. Vários pares de quasares foram observados tendo *redshifts* virtualmente idênticos e aparecem separados por pequenos intervalos angulares: esses pares são interpretados como sendo imagens gêmeas de um mesmo quasar. No livro de Kenyon[7] (pág.99) é mostrada uma inequívoca evidência para essa interpretação obtida no caso dos quasares UM6673A e UM673B que de fato são imagens gêmeas de um mesmo quasar.

     O efeito de lente gravitacional é um forte indício de que a TGE é válida para todo o Universo e não só localmente para o Sistema Solar.

## 3. Atraso Temporal de Sinais Luminosos

     Um outro efeito observável é o do atraso temporal sofrido por um sinal de radar enviado da Terra para um planeta alvo que é refletido de volta para a Terra. Usualmente denominado de "atraso temporal de ecos de radar". Esse atraso é provocado pelo campo gravitacional do Sol que reduz a velocidade de propagação dos sinais luminosos.

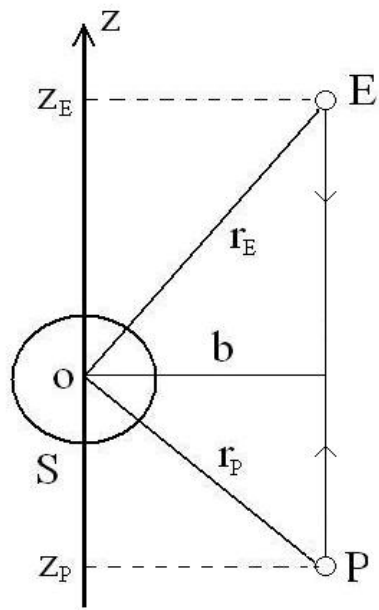

**Figura 2.** Trajetória do radar, ida (E→P) e volta (P→E), entre a Terra (E) e o planeta (P).

     O caminho, de ida E→P e volta P→E, seguido pelo raio de luz, ilustrado na Fig.2, é enviado da Terra (E) para o planeta alvo (P) passando

nas vizinhanças do Sol (S) a uma distância mínima b do mesmo. De acordo com (I.2) teremos,

$$ds^2 = [(1- \kappa/2r)/(1 + \kappa/2r)]^2 \, c^2 dt^2 - (1 + \kappa/2r)^4 \, (dx_1^2+dx_2^2+dx_3^2) \quad (3.1),$$

onde $\kappa = GM_S/c^2$ e $M_S$ é a massa do Sol. Desprezando termos da ordem de $(\kappa/r)^2$ em (3.1) obtemos:[4]

$$ds^2 = (1- 2\kappa/r) \, c^2 dt^2 - (1 + 2\kappa/r) \, (dx_1^2+dx_2^2+dx_3^2) \quad (3.2).$$

Como a luz descreve uma geodésica (ds = 0) verificamos que a velocidade v de propagação da luz ao longo trajetória medida, por exemplo, por um observador longe do campo gravitacional do Sol é dada por

$$v = d\xi/dt = [(dx_1^2+dx_2^2+dx_3^2)]^{1/2}/dt = c \, (1 - 2\kappa/r) \quad (3.2),$$

onde $d\xi$ é o elemento de comprimento de arco da trajetória. De acordo com (3.2) a velocidade da luz é sempre menor do que c. Note-se que o elemento de arco $d\xi$ e a velocidade v são medidos por um observador muito longe de um campo gravitacional. A velocidade da luz medida em um referencial local sobre qualquer ponto da trajetória é c. Assim, o tempo de percurso do sinal de radar emitido da Terra (no ponto E) até para o planeta alvo em P é dado por[8]

$$\Delta T_{EP} = \int_E^P d\xi/v = \int_E^P (1 + 2\kappa/r) \, d\xi/c \quad (3.3),$$

onde a integral é calculada ao longo da trajetória da luz que na Fig.1 está representada pela linha reta EP. Como vimos na Seção 2 a trajetória da luz é uma hipérbole, mas como o desvio da luz $\Delta\varphi = 4GM_S/R_S c^2$ é muito pequeno a sua trajetória pode ser aproximada por uma reta[8] (linha EP na Fig.1) com um erro desprezível da ordem de $\Delta\varphi^2$. Desse modo, fazendo $d\xi = dz$ e integrando (3.3) do ponto E até P com $r = (z^2 + b^2)^{1/2}$ [*vide comentários sobre distâncias astronômicas no* **Apêndice**] onde b é o ponto da trajetória mais próximo do Sol,

$$\Delta T_{EP} = \int_E^P (1 + 2\kappa/r) dz/c = (z_P - z_E)/c +$$
$$(2\kappa/c) \ln\{[z_P + (z_P^2 + b^2)^{1/2}]/[z_E + (z_E^2 + b^2)^{1/2}]\} \quad (3.4),$$

onde o parâmetro b não pode ser menor do que o raio do Sol. O primeiro termo de (3.4) dá o tempo de percurso num espaço-tempo plano e o segundo representa o tempo de percurso extra de atraso gerado pelo campo gravitacional do Sol. Note-se que $\Delta T_{EP}$ independe da freqüência da luz e pode-se mostrar[8] que a freqüência da luz não varia ao longo da geodésica.

O atraso maior ocorre quando a Terra e o planeta alvo estão em lados opostos do Sol, quase na "conjunção superior". Neste caso b/z << 1 e o tempo de atraso Δt de ida e volta ( Terra → planeta → Terra) é dado por

$$\Delta t = (4GM_S/c^3) \ln(4|z_E| z_P/b^2) \qquad (3.5).$$

Na "conjunção superior" temos $|z_1| \approx r_E$ = raio da órbita da Terra ≈ 14.9 $10^{10}$ m e $z_P \approx r_P$ = raio da órbita do planeta. Obviamente, b não pode ser menor do que o raio do $R_S$ do Sol. Para o caso do raio tangente ao Sol fazendo b = $R_S$ = 6.96 $10^8$ m verificamos que o tempo de atraso na conjunção superior dado por (3.5) para o planeta Vênus é ~220 µs. O tempo de percurso de ida e volta do sinal para Vênus no caso de um espaço plano $2(z_P - z_E)/c \sim 2(r_P - r_E)/c$ = 1300 µs.

Notemos que o tempo de atraso Δt medido na Terra deve ser dado por $\Delta\tau = \Delta t (1-2GM_S/r_E c^2)$, onde $r_E$ é o raio da órbita da Terra em torno do Sol. Ou seja, Δτ é o tempo próprio (local) medido na Terra. Outros fatores de correção são levados em conta tais como a mudança de velocidade da luz nos plasmas da coroa solar e interplanetário e a distorção gerada pela topografia das superfícies dos planetas que refletem o radar.[2,7,8] Atrasos de ecos de radar para Mercúrio,Vênus[19,20] e Marte[21] foram medidos por Shapiro et al. usando os radio-telescópios de Arecibo (Porto Rico) e Haystack (Massachusetts, USA). Após vários anos de medidas muito meticulosas e precisas verificou−se que os valores teóricos e experimentais concordam a menos de um fator 1.02 ± 0.02. Os sinais de radar no caso de Marte foram recebidos e refletidos pela sonda espacial "Viking Lander" pousada em Marte. Outras medidas de tempo de atraso de ecos de radar foram efetuadas por Anderson et al.[22] utilizando sinais das sondas espaciais Mariner 6 e 7 em órbita em torno do Sol. Nesses casos os acordos obtidos entre os resultados teóricos e experimentais são da ordem de 1.00 ± 0.03.

Enfim, essas medidas mostram um outro claro sucesso para a TGE.

## 4. Movimento Planetário no Sistema Solar.

Estudos recentes[1,2,7,23] foram feitos sobre os desvios da teoria Newtoniana descrevendo os movimentos planetários no Sistema Solar. Esses trabalhos envolveram milhares de medidas nas quais foram usadas modernas técnicas de observação. Através de "best fits" entre os resultados experimentais e as previsões dadas pelas equações de campo da TGE deveríamos esperar que o tensor métrico $g_{\mu\nu}$ obtido deveria ser dado, segundo (I.1), por $g_{oo} = (1- 2GM_S/rc^2)$ , $g_{ok} = -1$ e $g_{ij} = -1/(1- 2GM_S/rc^2)$. Verificou−se que

$$g_{oo} = (1- 2GM_S/rc^2) + \beta\, 2(GM_S/rc)^2 + \alpha\, (GM_S/r)(w/c)^2$$

$$g_{ok} = -\alpha\,(GM_S/r)(w^k/c) \quad \text{e} \quad g_{ij} = -\delta_{ij}/(1- 2\gamma GM_S/rc^2),$$

onde $w^k$ é a velocidade da Terra em relação a um referencial definido pela radiação cósmica[7] de fundo e os parâmetros $\alpha$, $\beta$ e $\gamma$ obtidos são dados respectivamente, por $\alpha = (2.2 \pm 1.8)\,10^{-4}$ e $\beta = 1 + (0.2 \pm 1.0)\,10^{-3}$ e $\gamma = 1 + (-1.2 \pm 1.6)\,10^{-3}$. que confirmam, mais uma vez, a consistência da TGE para explicar os movimentos planetários no Sistema Solar.

## 5. Conclusões.

De acordo com o que vimos neste artigo, a TGE explica com sucesso, dentro dos limites dos erros experimentais, a precessão dos periélios dos planetas, deflexão de luz pelo Sol, o retardo de ecos de sinais de radar passando ao redor do Sol e o movimento planetário no Sistema Solar.

**APÊNDICE. Distâncias Astronômicas Euclideanas.**

As equações que usamos para descrever os vários processos no Sistema Solar são funções de vários parâmetros tais como raio do Sol, distâncias entre planetas, distâncias entre planetas e o Sol, afélio, periélio, etc, que são medidos pelos astrônomos[24] assumindo uma geometria euclideana que é válida no limite Newtoniano. Entretanto, como sabemos, a geometria na qual os corpos celestes estão imersos, no contexto da TGE, é a 4−dim de Riemann. De acordo a TGE a distância entre dois pontos definidos, por exemplo, pelas coordenadas $r_1$ e $r_2$ não é igual $r_2 - r_1$ mas depende da curvatura do espaço−tempo[4−9]. Assim, quais são os erros cometidos aos assumirmos, no Sistema Solar, que as dimensões dos astros e que as entre distâncias entre eles são as astronômicas euclideanas?

Ora, de acordo com a métrica de Schwarzschild (I.1) a distância $\Delta\ell$ entre dois pontos, um na superfície do Sol com coordenada $r = R_S$ e outro com coordenada $r = R$ é dada por[4−9]

$$\Delta\ell = \int_{R_S}^{R} dr/(1 - 2\kappa/r)^{1/2} =$$

$$(R - R_S)\{1+ [\kappa/(R - R_S)]\ln(R/R_S) + (3/2)(\kappa^2/RR_S) + ...\} \quad (A.1),$$

onde $(R - R_S)$ é distância euclideana e $\kappa = GM_S/c^2 \approx 1.70\,10^3$ m. De acordo com (A.1) mesmo para pontos muito próximos da superfície do Sol, $R \approx 2.5\,R_S$ verificamos que o maior termo da expansão $\kappa/(R - R_S)]\ln(R/R_S) \sim (8\,10^6)^{-1}$. Isto mostra que usando uma geometria euclideana para medir distâncias no Sistema Solar introduzimos um erro na medida da distância que é inferior a uma parte em oito milhões que é muitíssimo menor do que os erros cometidos nas medidas astronômicas.[24]

# REFERÊNCIAS

[1] C.M.Will. "Theory and Experiment in Gravitational Physics" Cambridge University Press, Cambridge (1993) Rec.

[2] C.M.Will. "The Confrontation between General Relativity and Experiment" . http://relativity.livingreviews.org/

[3] M.Cattani. http://arxiv.org/abs/1005.4314 (2010).

[4] G.C.McVittie. "General Relativity and Cosmology", Chapman and Hall Ltd, London (1965).

[5] H.Yilmaz. "Theory of Relativity and the Principles of Modern Physics", Blaisdell Publishing Company, NY(1965).

[6] C.W.Misner, K.S.Thorne and J.A.Wheeler. "Gravitation", Freeman (1970).

[7] I.R.Kenyon. "General Relativity", Oxford University Press (1990).

[8] H.C.Ohanian. "Gravitation and Spacetime".W.W.Norton (1976).

[9] M.Cattani. http://arxiv.org/abs/1001.2518 ; http://arxiv.org/abs/1003.2105 ; thttp://arxiv.org/abs/1004.2470/ (2010).

[10] L.Landau et E.Lifchitz."Théorie du Champ",Éditions de la Paix(1964).

[11] M.Cattani. Revista Brasileira de Ensino de Física 20, 27 (1998).

[12] A.Sommerfeld. "Mechanics", University of California (1952)

[13] L.Landau e E.Lifchitz."Mecânica". Hemus–Livraria e Editora (SP) (1970).

[14] K.R.Symon. "Mechanics", Addison–Wesley (1957).

[15] http://searadaciencia.ufc.br/especiais/fisica/sobral1919/sobral7.htm[

[16] D.E.Lebach, B.E.Corey, I. I. Shapiro, M.I. Ratner, J.C. Webber, A. E. E. Rogers, J.L. Davis and T.A. Herring. Phys. Rev. Lett.**75**, 1439 (1995).

[17] M.Froeschle, F.Mignard and F.Arenou. *Proceedings of the Hipparcos Venice 1997 Symposium*, http://astro.estec.esa.nl/Hipparcos/venice.html / .

[18] T. Müller and D. Weiskopf. Am. J. Phys.78, 204 (2010).

[19] I.I.Shapíro, M.E.Ash, R.P.Ingalls,W.B.Smith, D.B.Campbell, R.B.Dyce, R.F.Jurgens and G.H.Pettengill. Phys.Rev.Lett.26,1132(1971).

[20] I.I.Shapiro, G.H.Pettengill, M.E.Ash, R.P.Ingalls, D.B.Campbell and R.B.Dyce. Phys. Rev. Lett. 28, 1594 (1978).

[21] R.D.Reasenberg, I.I.Shapiro, P.E.MacNeil, R.B.Goldstein, J.C.Breindenthal, J.P.Brenkle, D.L.Cain,T.M.Kaufman,T.A.Komarek and A.I.Zygielbaum. Ap.J. 234, L314–221(1979).

[22] J.D.Anderson, P.B.Esposito,W.Martin, C.L.Thornton and D.O.Muhleman. Ap.J. 200, 221 (1975).

[23] R.W.Hellings. "General relativity and gravitation"pp.365–388, Reidel, Dordrecht (1984).

[24] L.Motz and A.Duveen."Essentials of Astronomy", Columbia University Press (1971).